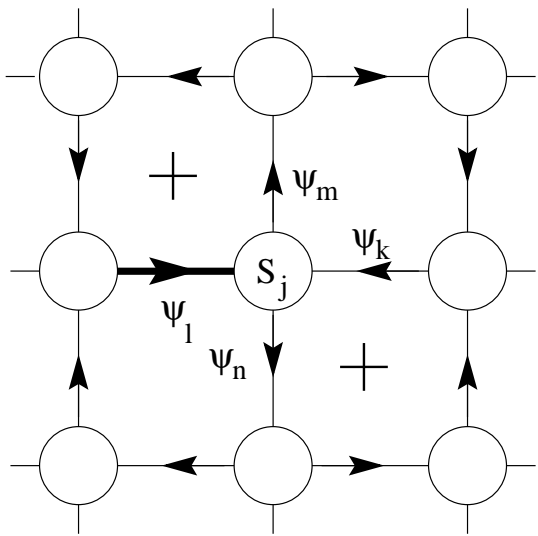 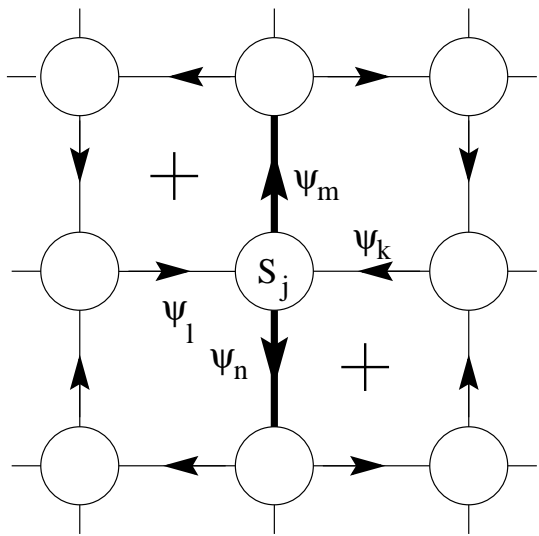

U

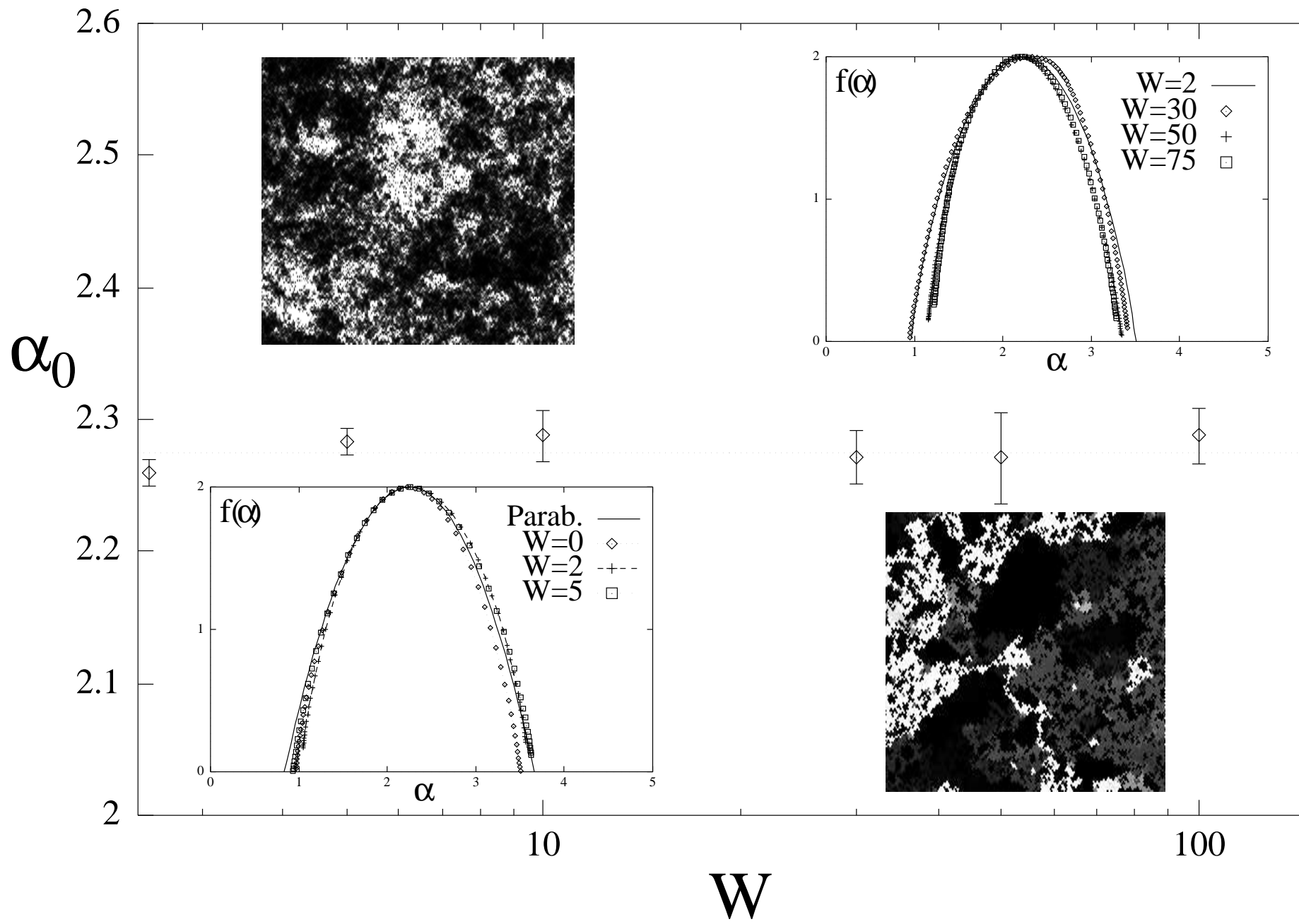

# Universal Multifractality in Quantum Hall Systems with Long-Range Disorder Potential [*]

Rochus Klesse and Marcus Metzler

*Institut für Theoretische Physik der Universität zu Köln*

*D-50937 Köln, Germany*

We investigate numerically the localization-delocalization transition in quantum Hall systems with long-range disorder potential with respect to multifractal properties. Wavefunctions at the transition energy are obtained within the framework of the generalized Chalker–Coddington network model. We determine the critical exponent $\alpha_0$ characterizing the scaling behavior of the local order parameter for systems with potential correlation length $d$ up to 12 magnetic lengths $l$. Our results show that $\alpha_0$ does not depend on the ratio $d/l$. With increasing $d/l$, effects due to classical percolation only cause an increase of the microscopic length scale, whereas the critical behavior on larger scales remains unchanged. This proves that systems with long-range disorder belong to the same universality class as those with short-range disorder.

Since the discovery of the Quantum Hall Effect (QHE), the localization-delocalization (LD) problem in two-dimensional (2D) disordered electron systems in a strong magnetic field has attracted much attention. It is now generally accepted that in the case of spinless non–interacting electrons only states in the center of the disorder broadened Landau bands are extended, while at all other energies the states are localized [1]. The role of fractality in this transition was already observed by Aoki [2] investigating inverse participation numbers. The inherent connection between localization transitions and *multifractality* was first recognized by Castelani and Peliti [3]. They have calculated generalized inverse participation numbers $(P(q) = \int |\psi(r)|^{2q} d^2 r)$ and found a non trivial $q$-dependence of the scaling behavior from which they concluded that wavefunctions at the critical point exhibit multifractal properties. Since then the multifractal analysis of the LD transition has been developed considerably [4]. In the case of the QHE, Pook, Janssen [5] and Huckestein, Schweitzer [6] have calculated the $f(\alpha)$-spectra of critical states and found them to be universal for quantum Hall systems with several types of short-range disorder potentials. Systems with long-range disorder, where the potential correlation length $d$ is large compared to the magnetic length $l$, have not yet been investigated, since the numerical effort increases rapidly with the ratio $d/l$ within the framework of the models that have been used up to now.

Indeed, the case of $d \gg l$ is especially interesting, because then a very appealing semi-classical picture applies in which the LD transition is connected to the 2D percolation transition [7]: In a smooth random potential $V$ and strong magnetic field a classical 2D-electron executes a fast cyclotron motion on a circle of radius $l$ around a guiding center, which drifts along a contour $V(r) = E'$. For this reason the probability density of eigenstates with energy $E = E' + \hbar\omega_c/2$ is concentrated on strips of width $l$ along these contours. An overlap between strips of separate contours at saddle points allows the electron to tunnel between them. Since this overlap decreases with decreasing $l$, tunneling becomes completely negligible in the limit $d/l \to \infty$. In this case the eigenstates are localized on single contours $V(r) = E'$ with critical size $\xi_p$ diverging $\propto |E' - E_c|^{\nu_p}$ as $E'$ approaches the percolation threshold $E_c$. Here $\nu_p = 4/3$ is the critical exponent belonging to the 2D-percolation transition. However, the LD-transition is – at least for small ratios $d/l$ – characterized by a correlation length diverging with a different critical exponent $\nu = 2.3$ [1]. So the question arises, whether at large (but finite) ratios $d/l$ effects due to classical percolation change the nature of the quantum mechanical LD-transition or not.

We investigate this problem by analyzing critical states of generalized Chalker-Coddington networks with respect to multifractality. These models have been introduced by Chalker, Coddington [9] and generalized by Chalker, Eastmond [10] and Lee et al. [11] to describe quantum Hall systems with long-range random potential and are therefore suitable for our purposes. We analyze critical wavefunctions of network models corresponding to ratios $d/l$ up to 12 and find that the characteristic multifractal exponent $\alpha_0$ is *independent* of $d/l$ and the same as for systems with short-range disorder investigated in [5,6]. So our numerical results directly confirm systems with $d/l$ varying from 0 up to 12 to be in the same universality class. Furthermore, a discussion of the influence of classical percolation effects leads us to the conclusion that this remains valid even for arbitrarily large (but finite) ratio $d/l$, if the system size is sufficiently large.

The network model consists of $2 \times 2$ unitary scattering

---

matrices representing saddle points of a smooth random potential which are arranged on a square lattice and connected by one-dimensional unidirectional channels, called links, corresponding to equipotential lines between the saddle points. The scattering matrices $S_j$ describe the transitions from electron states of incoming links $\psi_k, \psi_l$ to outgoing link states $\psi_m, \psi_n$ (see fig 1),

$$\begin{pmatrix} \psi_m \\ \psi_n \end{pmatrix} = S_j \begin{pmatrix} \psi_k \\ \psi_l \end{pmatrix}, \qquad S_j = \begin{pmatrix} t_{mk} & t_{ml} \\ t_{nk} & t_{nl} \end{pmatrix}. \quad (1)$$

Traversing a link $k$ an electron acquires an additional phase $\varphi_k$, randomly distributed but constant for each link, which we absorb in the complex scattering coefficients $t_{mk}, t_{nk}$. The transmission amplitudes $T_j = |t_{mk}|^2 = |t_{nl}|^2$, $R_j = |t_{ml}|^2 = |t_{nk}|^2$ are determined by the difference between electron energy $E$ and random saddle point energy $u_j \in [-W, W]$, $T_j^{-1} = 1 + \exp((u_j - E)/\Delta)$, $R_j = 1 - T_j$. The relevant parameter is the ratio of the width $W$ of the saddle point distribution and the tunneling energy $\Delta$; $W/\Delta = 0$ yields the original Chalker-Coddington model, whereas $W/\Delta \to \infty$ corresponds to classical percolation, since then the transmission amplitude is zero or unity at every saddle point.

To compare this model with explicit smooth disorder models we have to relate $W$ and $\Delta$ to the magnetic length $l$ and to the strength $V_0 = \langle V^2 \rangle^{1/2}$ and correlation length $d$ of the disorder potential. The energy distribution of the saddle points is limited by the potential strength $V_0$, therefore $W \approx V_0$. For a real saddle point of the random potential with curvature $c = |V_{xx} V_{yy}|^{1/2}$ the inverse transmission amplitude is $1 + \exp((u_j - E) 2\pi/cl^2)$ [12], so with the estimate $c \approx V_0/d^2$ one gets $\Delta \approx V_0(l/d)^2$ and $W/\Delta \approx (d/l)^2$.

A wavefunction $\Psi$ on the network is given by its complex amplitudes $\psi_l$ on each link $l$, i. e. $\Psi \in \mathcal{C}^N$, where $N$ is the number of links. It is stationary at energy $E$ if the scattering condition

$$\psi_m = t_{mk}\psi_k + t_{ml}\psi_l \quad (2)$$

is satisfied at each saddle point. For our purpose it is convenient to use the disorder and energy dependent unitary operator $U$ defined by

$$U e_l = t_{ml} e_m + t_{nl} e_n \quad (3)$$

(see fig. 1), where $e_j$ denotes the unity vector whose components vanish everywhere except at link $j$ [13]. Now condition (2) simply states

$$U\Psi = \Psi. \quad (4)$$

For a given disorder configuration this equation has solutions only for discrete energies. Instead of explicitly looking for these energies, which would be very cumbersome, we fix $E$ and solve the more general eigenvalue equation $U_E \Psi = e^{i\varphi} \Psi$. A solution $\Psi_\varphi$ of this equation is obviously a stationary wavefunction at energy $E$ for a disorder configuration corresponding to $U' := e^{-i\varphi}U$, i.e. a configuration that deviates from the initial one only by a simple global shift of the random phases. The latter transformation does not influence the randomness of the coefficients, hence it does not change the statistical properties of the wavefunctions we are interested in.

The square amplitudes of the normalized wavefunctions can be subjected to a multifractal analysis [4]: divide the system of size $L$ into $N = \lambda^{-2}$ boxes $b_i(\lambda)$ of size $L_b = \lambda L$, $\lambda \in [0, 1]$ and determine the scaling behavior of the averaged $q$-moments of the box-probabilities

$$\langle P^q(\lambda) \rangle = \frac{1}{N} \sum_i P_i^q, \quad P_i = \int_{b_i(\lambda)} d^2 r |\Psi(r)|^2, \quad (5)$$

with respect to $\lambda$ for real values of $q$. The scaling behavior can be described by a function $\tau(q)$ that is defined by

$$\langle P^q(\lambda) \rangle \propto \lambda^{D+\tau(q)}, \quad (6)$$

where in our case $D = 2$. $\tau(q)$ obviously increases monotonically with $q$ due to normalization and has – less obviously – negative curvature, see e.g. [4]. Therefore its Legendre transformed, the so called $f(\alpha)$-spectrum,

$$\alpha(q) = \frac{d\tau}{dq}, \quad f(\alpha(q)) = \alpha(q) \cdot q - \tau(q). \quad (7)$$

is a single-humped, positive valued function, restricted to $D_{+\infty} \le q \le D_{-\infty}$, where $D_{\pm\infty} = \alpha(q \to \pm\infty)$. The $f(\alpha)$-spectrum describes the scaling behavior of the entire distribution function of box probabilities and takes its maximum value $D$ at $\alpha_0 \equiv \alpha(q = 0)$. This value $\alpha_0$ is the critical exponent describing the *typical* scaling of the box-probabilities, $P_{typ} \equiv \exp\langle \ln P \rangle \propto \lambda^{\alpha_0}$. In phase-coherent systems the distribution of the box probabilities is of *log*-normal type corresponding to a parabolic $f(\alpha)$-spectrum [4]

$$f(\alpha) = -\frac{(\alpha - \alpha_0)^2}{4(\alpha_0 - D)} + D. \quad (8)$$

Indeed, this parabolic approximation, fitted only by $\alpha_0$, describes the $f(\alpha)$-spectrum obtained in [5,6] in a broad range around the maximum. For these reasons $\alpha_0$ is the most important scaling exponent and least influenced by statistical fluctuations. Other characteristic exponents are $D_{\pm\infty}$ describing the scaling of the minimum (-) and maximum (+) box-probabilities, $P_\pm \propto \lambda^{D_{\pm\infty}}$, and the exponent $\tau(2)$ of the inverse participation number. The latter is connected to the anomalous diffusion exponent $\eta$ [14] via the scaling relation $\eta = D - \tau(2)$ [5,8].

We analyze wavefunctions at the critical energy $E = 0$ for $\Delta = 1$ and $W$ varying from 0 to 200 for networks of $150 \times 150$ saddle points on a cylinder. As expected, the amplitude distributions of these wavefunctions are of log–normal type and a multifractal analysis can be applied.

For $W \lesssim 75$ we obtain almost the same $f(\alpha)$-spectra as for systems with short-range disorder potential (see fig.2). For $W \gtrsim 75$ the sample to sample fluctuations and hence the statistical error increases rapidly, especially for large values of $|q|$. In order to obtain proper $f(\alpha)$-spectra we had to choose larger minimal box sizes for increasing values of $W$. At $W \approx 200$ even for $q = 0$ this minimum box size approaches the system size, and thus for $W \gtrsim 150$ no $f(\alpha)$-spectra can be obtained.

The exponent $\alpha_0$ shows no significant dependence on the width $W$, even for the largest values (see fig. 2). The value averaged over all systems is $\alpha_0 = 2.27 \pm 0.01$. For the original Chalker-Coddington model ($W = 0$) we obtain $\alpha_0 = 2.26 \pm 0.01$. This is in good agreement with the values $2.30 \pm 0.07$ and $2.29 \pm 0.02$ obtained for critical states in short ranged disorder systems in [5] and [6], respectively. Furthermore, we find $D_{-\infty} = 3.6 \pm 0.2$, $D_{\infty} = 1.0 \pm 0.2$ and the anomalous diffusion exponent $\eta = 0.4 \pm 0.1$ also very close to those in [5] ($D_{-\infty} = 3.7 \pm 0.1$, $D_{\infty} = 0.95 \pm 0.1$) and [14,15] ($\eta = 0.38 \pm 0.04$). These results confirm the hypothesis of universal multifractality for long-range disorder potentials with correlation lengths up to $d \approx \sqrt{150}l \approx 12l$. In the following we argue that this scaling behaviour does not change in the limit $L \to \infty$ even for arbitrarily large ratios $W/\Delta$ and $d/l$.

With this end in mind we discuss the influence of classical percolation on this LD-transition. The original Chalker-Coddington model ($W = 0$) exhibits no classical percolation, because all saddle points are at the same energy, whereas in the generalized model percolation is incorporated and defines a new length scale. The microscopic length of the original model is simply given by the lattice constant $a$ corresponding to the potential correlation length $d$. For the generalized model with $W > \Delta$ the microscopic length is significantly enhanced by the occurrence of saddle points whose energies $u_j$ differ more than $\Delta$ from the electron energy $E = 0$. At these saddle points the coefficient $R_j$ (or $T_j$) is exponentially small ($\sim \exp(-|u_j|/\Delta)$) and the amplitude in an incoming link is essentially transmitted in a single outgoing link only (classical behavior). Thus the amplitude of the wavefunction remains constant on percolating paths avoiding saddle points with energy $|u_j| < \Delta$. The typical distance covered by these classical paths sets a new microscopic length scale as a function of $W/\Delta$, $a(W/\Delta)$. Figure 2 clearly shows the appearance of the new length scale: the areas of approximately constant amplitude are considerably larger in the grey-scale plot on the right than on the left corresponding to wavefunctions of systems with $W/\Delta = 150$ and $W/\Delta = 0$, respectively. Identifying these classical paths with links, a model with arbitrarily large $W$ can be rescaled to a model with lattice constant $a' = a(W/\Delta)$ and $W' \approx \Delta$ [16]. So the influence of classical percolation can be eliminated by renormalizing the lattice constant and universal behavior is to be expected as long as the system size is larger than $a(W/\Delta)$. This explains the above-mentioned necessity to choose larger minimum box sizes, since they have to be larger than the microscopic length, and the breakdown of the multifractal analysis when $a(W/\Delta)$ reaches the system size.

To summarize our findings, we gave numerical evidence that systems with potential correlation lengths up to 12 times the magnetic length have the same critical exponent $\alpha_0$ as short-range disorder systems. We also argued that $\alpha_0$ will not change even for much larger ratios of $d/l$ as long as the systems are sufficiently large. This is based on the observation that classical percolation which could be responsible for such a change only leads to an increase of the microscopical length scale that can be accounted for by rescaling the system. This leads us to the conclusion that the LD transition in quantum Hall systems is universal for disorder potentials with arbitrary correlation length.

FIG. 1. The Chalker-Coddington network. At each saddle point the scattering matrix $S_j$ describes the transition from incoming to outgoing states. The operator $U$ maps each incoming link amplitude to the two outgoing links.

FIG. 2. The $\alpha_0$ values for different saddle point distributions. ($W$ is plotted logarithmically.) The insets show the $f(\alpha)$–spectra of the wavefunctions and their amplitudes in a grey-scale plot, where darker areas denote lower square amplitude. On the left one can see spectra for $W = 0$ to 5 compared to the parabolic approximation and one wavefunction for $W = 0$ and on the right $W = 30$ to 75 with $W = 2$ as reference and the plot depicts a wavefunction for $W = 150$.